\documentstyle[onecolumn]{mn}
\newcommand{\mincir}{\raise
-2.truept\hbox{\rlap{\hbox{$\sim$}}\raise5.truept
\hbox{$<$}\ }}
\newcommand{\magcir}{\raise
-2.truept\hbox{\rlap{\hbox{$\sim$}}\raise5.truept
\hbox{$>$}\ }}
\newcommand{\minmag}{\raise-2.truept\hbox{\rlap{\hbox{$<$}}\raise
6.truept\hbox
{$>$}\ }}
\newcommand{\be}{\begin{equation}}
\newcommand{\ee}{\end{equation}}
\newcommand{\ba}{\begin{eqnarray}}
\newcommand{\ea}{\end{eqnarray}}
\newcommand{\brr}{\begin{array}}
 
\newcommand{\err}{\end{array}}
\newcommand{\bc}{\begin{center}}
\newcommand{\ec}{\end{center}}

\input{psfig.sty} 
\title{The correlation function of X-ray galaxy clusters in the
RASS1 Bright Sample}
\author[Moscardini et al.]
{L. Moscardini$^1$, S. Matarrese$^{2,*}$,
S. De Grandi$^3$  and F. Lucchin$^1$ \\
$^1$Dipartimento di Astronomia, Universit\`a di Padova,
vicolo dell'Osservatorio 5, I--35122 Padova, Italy\\
$^2$Dipartimento di Fisica G. Galilei, 
Universit\`{a} di Padova, via Marzolo 8, I--35131 Padova, Italy\\
$^3$Osservatorio Astronomico di Brera, via Bianchi 46, 
I--23807 Merate (LC), Italy\\
$^*$Present address: Max-Planck-Institut f\"ur Astrophysik, 
Karl-Schwarzschild-Strasse 1, D-85748 Garching, Germany}

\begin{document}

\maketitle

\begin{abstract}
We analyse the spatial clustering properties of the RASS1 Bright
Sample, an X-ray flux-limited catalogue of galaxy clusters selected
from the southern part of the $ROSAT$ All-Sky Survey. The two-point
correlation function $\xi(r)$ of the whole sample is well fitted (in
an Einstein-de Sitter model) by the power-law $\xi=(r/r_0)^{-\gamma}$,
with $r_0= 21.5^{+3.4}_{-4.4}\ h^{-1}$ Mpc and
$\gamma=2.11^{+0.53}_{-0.56}$ (95.4 per cent confidence level with one
fitting parameter). We use the RASS1 Bright Sample as a first
application of a theoretical model which aims at predicting the
clustering properties of X-ray clusters in flux-limited surveys for
different cosmological scenarios. The model uses the theoretical and
empirical relations between mass, temperature and X-ray cluster
luminosity, and fully accounts for the redshift evolution of the
underlying dark matter clustering and cluster bias factor. The
comparison between observational results and theoretical predictions
shows that the Einstein-de Sitter models display too low a correlation
length, while models with a matter density parameter $\Omega_{\rm
0m}=0.3$ (with or without a cosmological constant) are successful in
reproducing the observed clustering. The dependence of the correlation
length $r_0$ on the X-ray limiting flux and luminosity of the sample
is generally consistent with the predictions of all our models.
Quantitative agreement is however only reached for $\Omega_{\rm 0m} =
0.3$ models. The model presented here can be reliably applied to
future deeper X-ray cluster surveys: the study of their clustering
properties will provide a useful complementary tool to the traditional
cluster abundance analyses to constrain the cosmological parameters.
\end{abstract}


\section{Introduction}
Galaxy clusters play an important role in the models for structure
formation based on the gravitational instability hypothesis. They are
the most extended gravitationally bound systems in the Universe. For
this reason the study of their properties is a useful tool to 
constrain the parameters entering in the definition of the
cosmological scenarios. In particular, their abundance and spatial
distribution (also as a function of redshift) have been used to obtain
estimates of the mass fluctuation amplitude and of the density parameter
$\Omega_0$ (e.g. Eke, Cole \& Frenk 1996; Viana \& Liddle 1996; Mo,
Jing \& White 1996; Oukbir, Bartlett \& Blanchard 1997; Eke et
al. 1998; Sadat, Blanchard \& Oubkir 1998; Viana \& Liddle 1999;
Borgani, Plionis \& Kolokotronis 1999;  Borgani et al. 1999). In the
past years many different groups have compiled deep cluster surveys
in the optical band, which have been used to compute the clustering
properties of galaxy clusters. The first results showed that clusters
are strongly correlated, with a correlation length $r_0\approx 20-25\
h^{-1}$ Mpc, a factor 4-5 larger than that obtained for local galaxies
(e.g. Bahcall \& Soneira 1983; Postman, Huchra \& Geller 1992).  Here
$h$ represents the Hubble constant $H_0$ in units of 100 km s$^{-1}$
Mpc$^{-1}$. Sutherland (1988) suggested the existence of a possible
strong effect due to the spurious presence of galaxies acting as
interlopers (see also Dekel et al. 1989; van Haarlem, Frenk \& White
1997). New analyses of optical catalogues, taking into account this
projection effect (Dalton et al. 1992; Nichol et al. 1992; Dalton et
al. 1994; Croft et al. 1997), led to a smaller value for the cluster
correlation length ($r_0\approx 13-18\ h^{-1}$ Mpc).

A way to overcome the projection problem is the use of data obtained
in the X-ray region of the spectrum. In fact, in this band, galaxy
clusters have a strong emission produced by thermal bremsstrahlung,
which allows to detect them also at high redshifts. Starting from the
eighties, different space missions produced extended cluster
catalogues which have been mainly used to compute their X-ray
luminosity function. In particular, the $ROSAT$ satellite provided a
good opportunity to build a reliable all-sky survey, which was
performed in the soft (0.1 -- 2.4 keV) X-ray band.  First studies of
the clustering properties in small samples of X-ray selected galaxy
clusters have been performed by Lahav et al. (1989), Nichol, Briel \&
Henry (1994) and Romer et al. (1994).  More recently the $ROSAT$ data
have been correlated with the Abell-ACO cluster catalogue (Abell,
Corwin \& Olowin 1989) to produce the X-ray Brightest Abell Cluster
sample (XBACs; Ebeling et al. 1996), for which estimates of the
two-point correlation function have been recently obtained (Abadi,
Lambas \& Muriel 1998; Borgani, Plionis \& Kolokotronis 1999). The
corresponding values for $r_0$ are in the range $r_0\approx 20-26\
h^{-1}$ Mpc. A smaller amplitude of the correlation function is
obtained from the preliminary analyses of the REFLEX sample (Collins
et al. 1999), which is also obtained by the $ROSAT$ All-Sky Survey
(RASS) data.

In this paper we estimate the clustering properties for the RASS1
Bright Sample (De Grandi et al. 1999), which is another X-ray cluster
catalogue obtained from the RASS. In this case the clusters are
spectroscopically searched in a preliminary list of candidates
produced by correlating the X-ray data with regions of galaxy
overdensity in the southern sky. In this way, the resulting catalogue
is not affected by the selection biases present in the Abell-ACO
cluster catalogue. The RASS1 Bright Sample is used to test a
theoretical model for the correlation function of X-ray clusters in
flux-limited samples (see also Moscardini et al. 1999). This model
makes use of the technique introduced by Matarrese et al. (1997) and
Moscardini et al. (1998), which allows a detailed modelling of the
redshift evolution of clustering, accounting both for the non-linear
dynamics of the dark matter distribution and for the redshift
evolution of the bias factor.  A characteristic feature of this
technique is that it takes into full account light-cone effects, which
are relevant in analysing the clustering of even moderate redshift
objects (see also Matsubara, Suto \& Szapudi 1997; de Laix \& Starkman
1998; Yamamoto \& Suto 1999).

The plan of the paper is as follows. In Section 2 we summarize the
characteristics of the RASS1 Bright Sample used in the following
clustering analysis. In Section 3 we discuss the method used to
compute the observational two-point correlation function for the RASS1
Bright Sample and we present the results. In Section 4 we introduce
our theoretical model to estimate the correlations of the X-ray
clusters in the framework of different cosmological models and we
compare our predictions to the observational results. 
Conclusions are drawn in Section 5.

\section {The Sample}
\begin{figure*}
\centering  
\psfig{figure=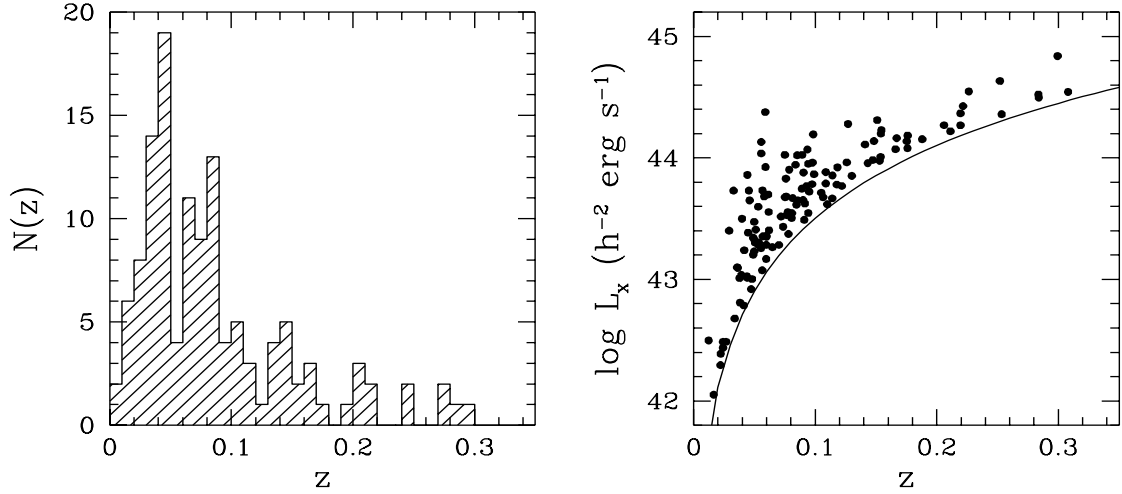,height=8.cm,width=16cm,angle=0}
\caption{Left panel: the redshift distribution of the RASS1 Bright Sample. 
Right panel: the X-ray luminosity $L_X$ (in the $ROSAT$ hard band 0.5
-- 2.0 keV) vs. the redshift $z$ for each cluster of the sample. The
solid line shows the X-ray luminosity corresponding to the
limiting flux $S_{\rm lim}= 3\times 10^{-12}$ erg s$^{-1}$ cm$^{-2}$
in an Einstein-de Sitter model.}
\label{fi:nz}
\end{figure*}

The RASS1 Bright Sample (De Grandi et al. 1999), contains 130 clusters
of galaxies selected from the first processing of the $ROSAT$ All-Sky
Survey (RASS) data (Voges 1992). This sample was constructed as part
of an ESO Key Programme (Guzzo et al. 1995) aimed at surveying all
southern RASS candidates, which is now known as the REFLEX cluster
survey (B\"ohringer et al. 1998; Guzzo et al. 1999). The
identification of RASS cluster candidates was performed by means of
different optically and X-ray based methods. First, candidates were
found as overdensities in the galaxy density distribution at the
position of the X-ray sources using the COSMOS optical object
catalogue (e.g. Heydon-Dumbleton, Collins \& MacGillivray 1989). Then,
correlating all RASS sources with the ACO cluster catalogue, and,
finally, selecting all RASS X-ray extended sources. X-ray fluxes were
remeasured using the steepness ratio technique (De Grandi et
al. 1997), specifically developed for estimating fluxes from both
extended and pointlike objects. A number of selections aimed at
improving the completeness of the final sample lead to the RASS1
Bright Sample. Considering the intrinsic biases and incompletenesses
introduced by the X-ray flux selection and source identification
processes, the overall completeness of the sample is estimated to be
$\magcir 90$ per cent.  The RASS1 Bright Sample is count-rate-limited
in the $ROSAT$ hard band (0.5 -- 2.0 keV), so that due to the
distribution of Galactic absorption its effective flux limit varies
between 3.05 and $4\times 10^{-12}$ erg cm$^{-2}$ s$^{-1}$ over the
selected area. This covers a region of approximately 2.5 sr within the
Southern Galactic Cap, i.e. $\delta<2.5^o$ and $b<-20^o$, with the
exclusion of patches with RASS exposure times lower than 150 s and of
the Magellanic Clouds area. The exact sky map covered by the sample is
shown in Figure 2 of De Grandi et al. (1999). The redshift
distribution for our whole sample is presented in the left panel of
Figure~\ref{fi:nz} while the X-ray luminosity $L_X$ as a function of
the redshift for each cluster is shown in the right one. It is
possible to notice that 66 per cent of the clusters have $z<0.1$ but
the redshift distribution has a tail up to $z\simeq 0.3$.
%--------------------------------------------------------
%--------------------------------------------------------
%

\section{The 2-point correlation function}
Before computing the clustering properties of our sample, we have to
derive for each cluster the comoving radial distance $r$ from the
observer, given the redshift of each source. To this goal we use the
standard relation (neglecting the effect of peculiar motions)
\be
r(z) =  {{c} \over H_0 \sqrt{|\Omega_{0\cal R}|}}     
  {\cal S} \left(\sqrt{|\Omega_{0\cal R}|}
\int_0^z dz' \left[\left( 1+z' \right)^2 \left(1+\Omega_{\rm 0m} z'\right) -
z'\left(2+z'\right) \Omega_{0\Lambda}\right]^{-1/2} \right)   \;, 
\label{eq:x_z}
\ee
where $\Omega_{0\cal R} \equiv 1 - \Omega_{\rm 0m} -
\Omega_{0\Lambda}$, with $\Omega_{\rm 0m}$ and $\Omega_{0\Lambda}$ the
density parameters for the non-relativistic matter and cosmological
constant components, respectively. In this formula, for an open
universe model, $\Omega_{0\cal R}>0$ and ${\cal S}(x)\equiv \sinh (x)$,
for a closed universe, $\Omega_{0\cal R}<0$ and ${\cal S}(x)\equiv \sin
(x)$, while in the Einstein-de Sitter (EdS) case, $\Omega_{0\cal R} =
0$ and ${\cal S}(x) \equiv x$. 

To compute the spatial two-point correlation function $\xi(r)$ we
adopt both the Landy \& Szalay (1993) estimator,
\be
\xi(r)= {{N_r (N_r-1)}\over{N_c (N_c-1)}}
 {{DD(r)}\over{RR(r)}} - {{N_r-1}\over {N_c}}
 {{DR(r)}\over{RR(r)}}+1\ ,
\label{eq:xi_ls}
\ee
and the Davis \& Peebles (1983) estimator, 
\be
\xi(r)= 2 {{N_r}\over {N_c-1}}  {{DD(r)}\over{RR(r)}}-1 \ .
\label{eq:xi_dp}
\ee
In the previous formulas $N_r$ is the number of 
random points and $N_c$ that of clusters, DD is the number of distinct
cluster-cluster pairs, DR is the number of cluster-random pairs and RR
refers to random-random pairs with separation between $r$ and $r +
\Delta r$. The random catalogue contains a number of sources 1,000 times 
larger than the real catalogue (i.e. $N_r=1000 N_c$).  To generate
this sample we have extracted randomly coordinates from the surveyed
area (see Figure 2 in De Grandi et al. 1999), assigning to each
position a random flux drawn from the observed number counts (Figure 8
in De Grandi et al. 1999).  We decided to retain the source in the
catalogue if its flux is larger than the flux limit at the choosen
position.  We adopt two different methods to assign the random
redshifts: in the first we scramble the observed redshifts of the
clusters in the sample; in the second we generate them randomly from
the observed redshift distribution binned in intervals of 0.01 in
$z$. The results obtained with these two different methods are
practically indistinguishable. The same happens also if we use the two
previous estimators for the two-point correlation function
(eqs.\ref{eq:xi_ls}-\ref{eq:xi_dp}).  The errorbars have been
estimated by using the bootstrap method with 50 resamplings. We find
that the errors obtained in this way are in many cases larger than
$\sqrt{3}$ times the Poissonian estimates, which are often used as an
analytical approximation of the bootstrap errors (Mo, Jing \& B\"orner
1992).  This is particularly true at small separations.

\begin{figure*}
\centering  
\psfig{figure=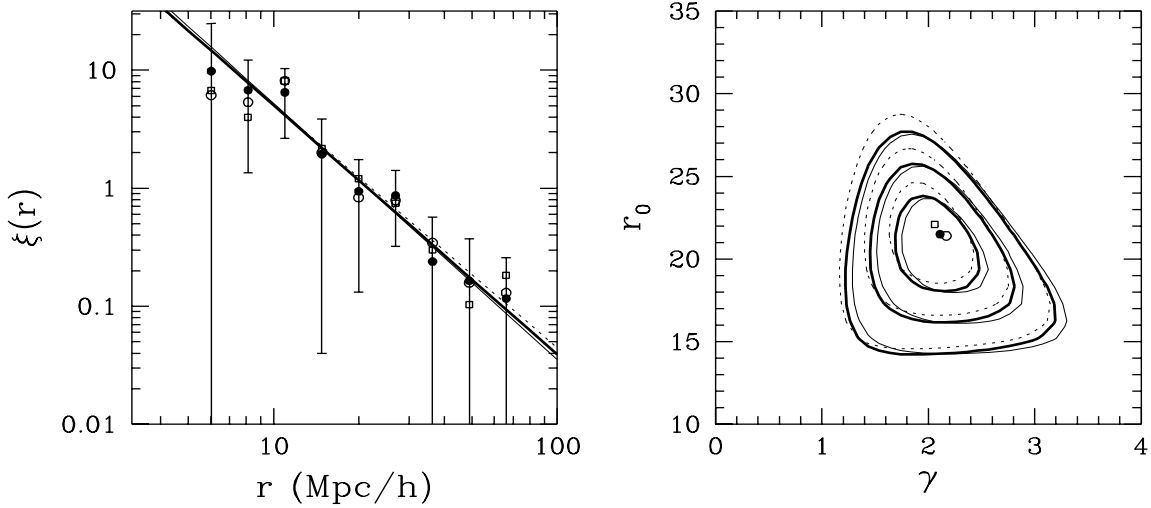,height=8.cm,width=16cm,angle=0}
\caption{Left panel: the two-point correlation function of the whole RASS1 
Bright Sample. Filled circles, open circles and open squares refer to
results obtained by assuming $(\Omega_{\rm
0m},\Omega_{0\Lambda})=(1,0)$, $(\Omega_{\rm
0m},\Omega_{0\Lambda})=(0.3,0)$ and $(\Omega_{\rm
0m},\Omega_{0\Lambda})=(0.3,0.7)$, respectively. The corresponding
best-fit relations are shown by heavy solid line, light solid line and
dotted line, respectively. The 1-$\sigma$ errorbars obtained by the
bootstrap method are shown only for the EdS case for clarity; their
sizes in the other cases are similar. Right panel: confidence contours
(68.3, 95.4  and 99.73 per cent confidence levels) of the fitting parameters
$r_0$ and $\gamma$ for the two-point correlation function of the whole
RASS1 Bright Sample. Heavy solid lines and filled circle refer to
confidence levels and to the best fitting value obtained for the EdS
case; light solid lines and open circle show the corresponding results
for $(\Omega_{\rm 0m},\Omega_{0\Lambda})=(0.3,0)$ while dotted lines
and open square refer to $(\Omega_{\rm
0m},\Omega_{0\Lambda})=(0.3,0.7)$ model.}
\label{fi:xitot}
\end{figure*}

In the left panel of Figure~\ref{fi:xitot} we show the correlation
function computed for the whole catalogue. We present results obtained
by using both an EdS model and two models with $\Omega_{\rm 0m}=0.3$
(with and without cosmological constant). The results are quite
similar and the differences are always small.

%--------------------------------------------------------
%--------------------------------------------------------
%

The correlation function has been fitted by adopting the power-law relation
\be
\xi(r)=(r/r_0)^{-\gamma}\ .
\label{eq:powlaw}
\ee
The best-fit parameters have been obtained by using a maximum
likelihood estimator based on Poisson statistics and unbinned data
(Croft et al. 1997; see also Borgani, Plionis \& Kolokotronis 1999).
Unlike the usual $\chi^2$-minimization, this method allows to avoid
the uncertainties due to the binsize, to the position of the bin
centres and to the bin scale (linear or logarithmic).

To build the estimator, it is necessary to estimate the predicted
probability distribution of cluster pairs, given a choice for the
correlation length $r_0$ and the slope $\gamma$.  By using all the
distances between the cluster-random pairs, we can compute the number
of pairs $g(r)dr$ in arbitrarily small bins $dr$ and use it to predict
the mean number of cluster-cluster pairs $h(r)dr$ in that interval as
\be
h(r)dr= 
{{N_c-1}\over{2N_r}} [1+\xi(r)] g(r)dr\ ,
\ee
where the correlation function $\xi$ is
modelled with a power-law as in eq.(\ref{eq:powlaw}).  Actually the
previous equation holds only for the Davis \& Peebles (1983) estimator
[eq.(\ref{eq:xi_dp})] but, since we obtain very similar results using
different estimators, we can safely apply it here. Now it is possible to
use all the distances between the $N_p$ cluster-cluster pairs data to
build a likelihood.  In particular, the likelihood function ${\cal L}$
is defined as the product of the probabilities of having exactly one
pair at each of the intervals $dr$ occupied by the cluster-cluster
pairs data and the probability of having no pairs in all the other
intervals.  Assuming a Poisson distribution, one finds
\be
{\cal L}= \prod_i^{N_p} \exp[-h(r)dr] h(r)dr \prod_{j\ne i} \exp[-h(r)dr]\ ,
\ee
where $j$ runs over all the intervals $dr$ where there are no pairs.
It is convenient to define the usual quantity $S=-2 \ln {\cal L}$
which can be written, once we retain only the terms depending on the
model parameters $r_0$ and $\gamma$, as
\be
S=2\int^{r_{\rm max}}_{r_{\rm min}} h(r)dr -2\sum_i^{N_p} \ln[h(r_i)]\ .
\ee
The integral in the previous equation is computed over the range of scales
where the fit is made.  We will adopt 5 and 80 $h^{-1}$ Mpc for
$r_{\rm min}$ and $r_{\rm max}$, respectively.

By minimizing $S$ one can obtain the best-fitting parameters $r_0$ and
$\gamma$; the confidence levels are defined by computing the increase
$\Delta_S$ with respect the minimum value of $S$ and assuming a
$\chi^2$ distribution for $\Delta_S$.

By applying this maximum likelihood method to the RASS1 Bright Sample
with the assumption of an EdS model, we find $r_0= 21.5^{+3.4}_{-4.4}\
h^{-1}$ Mpc and $\gamma=2.11^{+0.53}_{-0.56}$ (95.4 per cent
confidence level with one fitting parameter). Since the redshift
distribution is shallow, the values obtained in other cosmologies are
quite similar: for $(\Omega_{\rm 0m},\Omega_{0\Lambda})=(0.3,0)$ we
find $r_0= 21.4^{+3.4}_{-4.6} \ h^{-1}$ Mpc and
$\gamma=2.17^{+0.55}_{-0.56}$, while for $(\Omega_{\rm
0m},\Omega_{0\Lambda})=(0.3,0.7)$ we find $r_0= 22.1^{+3.6}_{-4.7} \
h^{-1}$ Mpc and $\gamma=2.06^{+0.54}_{-0.56}$. The best-fit relations
are also shown in Figure~\ref{fi:xitot}.  Notice that a
$\chi^2$-minimization procedure gives similar results, but with larger
errorbars.

In the right panel of the same figure we show the contour levels
corresponding to $\Delta_S$ equal to 2.30, 6.31 and 11.8.  Assuming
that $\Delta_S$ is distributed as a $\chi^2$ distribution with two
degrees of freedom, they correspond to 68.3, 95.4 and 99.73 per cent
confidence levels, respectively.  Notice that by assuming a Poisson
distribution the method considers all pairs as independent, neglecting
their clustering. Consequently the resulting errobars can be
underestimated (see the discussion in Croft et al. 1997).

Our results are somewhat larger than those derived by Romer et
al. (1994) who found $r_0 = 13-15\ h^{-1}$ Mpc by analysing a sample
of galaxy clusters also selected from the RASS in a similar region of
sky ($22^h <$ RA $< 3^h$, $-50^o < \delta < 2^o$, $|b|>-40^o$).  A
partial explanation of this difference is related to the deeper
limiting flux ($S_{\rm lim}\simeq 10^{-12}$ erg s$^{-1}$ cm$^{-2}$ in
the 0.1 -- 2.4 keV band) of their catalogue: as we will discuss in the
next section, the correlation length is expected to depend on the
characteristics defining the surveys, such as their limits in flux
and/or luminosity.  However we have to remind that this early sample
was derived drawing on X-ray information from the $ROSAT$ standard
analysis software (SASS), which was not optimazed for the analysis of
extended sources (for a more detailed discussion see e.g. De Grandi et
al. 1997), and this source of incompleteness was not included in the
analysis of Romer et al. (1994). Moreover in their analysis the sample
sky coverage (i.e., the surveyed area as a function of the flux limit)
was not discussed.

Previous analyses of the XBACs sample, which is a flux-limited
catalogue of X-ray Abell clusters with a limiting flux $S_{\rm lim}=
5\times 10^{-12}$ erg s$^{-1}$ cm$^{-2}$ in the 0.1 -- 2.4 keV band,
gave compatible amplitudes for the correlation function:
$r_0=21.1^{+1.6}_{-2.3}\ h^{-1}$ Mpc (Abadi, Lambas \& Muriel 1998)
and $r_0=26.0^{+4.1}_{-4.7}\ h^{-1}$ Mpc (Borgani, Plionis \&
Kolokotronis 1999; errorbars in this case are 2-$\sigma$
uncertainties). Preliminary analyses of the clustering properties of
the REFLEX sample (Collins et al. 1999; Guzzo et al. 1999), which has
a limiting flux $S_{\rm lim}= 3\times 10^{-12}$ erg s$^{-1}$ cm$^{-2}$
in the 0.1 -- 2.4 keV band, lead to a smaller correlation length
($r_0\simeq 18 \ h^{-1}$ Mpc). Also in this case, the discrepancy is
probably a consequence of the deeper limiting flux.
  
\section{Comparison with theoretical models}

\subsection{Structure formation models}

In the following analysis we consider five models, all normalized to
reproduce the local cluster abundance. In particular we will adopt the
normalizations obtained by Eke, Cole \& Frenk (1996) by analysing the
temperature distribution of X-ray clusters (Henry \& Arnaud 1991).
All our models belong to the general class of Cold Dark Matter (CDM)
scemarios; their linear power-spectrum can be represented as $P_{\rm
lin}(k,0) \propto k^n T^2(k)$, where, for the CDM transfer function
$T(k)$, we use the Bardeen et al. (1986) fit. In particular, we
consider three different EdS models, for which the power-spectrum
amplitude corresponds to $\sigma_8=0.52$ (here $\sigma_8$ is the
r.m.s. fluctuation amplitude in a sphere of $8 h^{-1}$ Mpc). They are:
a version of the standard CDM (SCDM) model with shape parameter (see
its definition in Sugiyama 1995) $\Gamma=0.45$ and spectral index
$n=1$; the so-called $\tau$CDM model, with $\Gamma=0.21$ and $n=1$; a
tilted model (TCDM), with $n=0.8$ and $\Gamma=0.41$, corresponding to
a high (10 per cent) baryonic content. We also consider an open CDM
model (OCDM), with matter density parameter $\Omega_{\rm 0m}=0.3$ and
$\sigma_8=0.87$ and a low-density flat CDM model ($\Lambda$CDM), with
$\Omega_{\rm 0m}=0.3$, with $\sigma_8=0.93$. Except for SCDM, which is
shown as a reference model, all these models are also consistent with
COBE data; for TCDM consistency is achieved by taking into account the
possible contribution of gravitational waves to large-angle CMB
anisotropies.  A summary of the parameters of the cosmological models
used in this paper is given in Table \ref{t:models}.

\begin{table}
\centering
\caption[]{The parameters of the cosmological models. Column 2: the present
matter density parameter $\Omega_{\rm 0m}$; Column 3: the present
cosmological constant contribution to the density $\Omega_{0\Lambda}$;
Column 4: the primordial spectral index $n$; Column 5: the Hubble
parameter $h$; Column 6: the shape parameter $\Gamma$; Column 7: the
spectrum normalization $\sigma_8$; Column 8: the value of the parameter $\eta$ 
in the luminosity-temperature relation required to reproduced the observed
$\log N$--$\log S$ (see text for details).}
\tabcolsep 4pt
\begin{tabular}{lccccccc} \\ \\ \hline \hline
Model & $\Omega_{\rm 0m}$ & $\Omega_{0\Lambda}$ & $n$ & $h$ &
$\Gamma$ & $\sigma_8$ & $\eta$  \\ \hline
SCDM         & 1.0 & 0.0 & 1.0 & 0.50 & 0.45 & 0.52 & -0.8 \\
$\tau$CDM    & 1.0 & 0.0 & 1.0 & 0.50 & 0.21 & 0.52 &  0.0 \\
TCDM         & 1.0 & 0.0 & 0.8 & 0.50 & 0.41 & 0.52 & -0.3 \\
OCDM         & 0.3 & 0.0 & 1.0 & 0.65 & 0.21 & 0.87 & -0.3 \\
$\Lambda$CDM & 0.3 & 0.7 & 1.0 & 0.65 & 0.21 & 0.93 & -0.2 \\
\hline
\end{tabular}
\label{t:models}
\end{table}

\subsection{The method}

Theoretical predictions for the spatial two-point correlation
function in the RASS1 Bright Sample have been here obtained in the
framework of the above cosmological models by using a method presented
in more detail in (Moscardini et al. 1999). 
Here we only give a short description.

Matarrese et al. (1997; see also Moscardini et al. 1998) developed an
algorithm to describe clustering in our past light-cone, where the
non-linear dynamics of the dark matter distribution and the redshift
evolution of the bias factor are taken into account. In the present
paper we adopt a more refined formula which better accounts for the
light-cone effects (see Moscardini et al. 1999). The observed spatial
correlation function $\xi_{\rm obs}$ in a given redshift interval
${\cal Z}$ is given by the exact expression
\be 
\xi_{\rm obs}(r) = { \int_{\cal Z} d z_1 d z_2 {\cal N}(z_1) r(z_1)^{-1} 
{\cal N}(z_2) r(z_2)^{-1} ~\xi_{\rm obj}(r;z_1,z_2) \over 
\bigl[ \int_{\cal Z} d z_1 {\cal N}(z_1) r(z_1)^{-1} \bigr]^2 } \;,
\label{eq:xifund}
\ee 
where $\xi_{\rm obj}(r,z_1,z_2)$ is the correlation function of pairs
of objects at redshifts $z_1$ and $z_2$ with comoving separation
$r$ and ${\cal N}(z)$ is the
actual redshift distribution of the catalogue. A related
approach to the study of correlations on the light-cone hypersurface
has been recently presented by Yamamoto \& Suto (1999) and Nishioka \&
Yamamoto (1999) within linear theory and by Matsubara, Suto \& Szapudi
(1997) in the non-linear regime.
 
An accurate approximation for $\xi_{\rm obj}$ over the scales
considered here is
\be 
\xi_{\rm obj}(r,z_1,z_2) \approx b_{\rm eff}(z_1) b_{\rm eff}(z_2) 
\xi_{\rm m}(r,z_{\rm ave}) \;,
\ee
where $\xi_m$ is the dark matter covariance function and $z_{\rm ave}$
is an intermediate redshift between $z_1$ and $z_2$, for which an
excellent approximation is obtained through $D_+(z_{\rm ave}) =
D_+(z_1)^{1/2}D_+(z_2)^{1/2}$ (Porciani 1997), with $D_+(z)$ the
linear growth factor of density fluctuations.

In our treatment we disregard the effect of redshift-space
distortions.  Some analytical expressions have been obtained in the
mildly non-linear regime, by using either the Zel'dovich approximation
(Fisher \& Nusser 1996) or higher order perturbation theory (Heavens,
Matarrese \& Verde 1998). The complicating role of the
cosmological redshift-space distortions on the evolution of the bias
factor has been considered by Suto et al. (1999).
A rough estimate of the effect of
redshift-space distortions can be obtained within linear theory and
the distant-observer approximation (Kaiser 1987; see Zaroubi \&
Hoffman 1996 for an extension of this formalism to all-sky
surveys). In this case the enhancement of the redshift-space averaged
power spectrum is given by the factor $1+2\beta/3+\beta^2/5$, where
$\beta\simeq \Omega_{\rm 0m}^{0.6}/b_{\rm eff}$ and $b_{\rm eff}$ is
the effective bias (see below). Plionis \& Kolokotronis (1998),
by analysing the XBACs catalogue and using linear perturbation theory
to relate the X-ray cluster dipole to the Local Group peculiar
velocity, found $\beta\simeq 0.24\pm 0.05$. Adopting this approach,
Borgani, Plionis \& Kolokotronis (1999) conclude that the overall
effect of redshift-space distortions is a small change of the correlation
function, which expressed in terms of $r_0$ corresponds to an  
$\simeq 8$ per cent increase.  
  
The effective bias $b_{\rm eff}$ appearing in the previous equation
can be expressed as a weighted average of the `monochromatic' bias
factor $b(M,z)$ of objects of some given intrinsic property $M$ (like
mass, luminosity, ...), as follows
\be 
b_{\rm eff}(z) \equiv {\cal N}(z)^{-1} \int_{\cal M} d\ln M' ~b(M',z) 
~{\cal N}(z,M')\, ,
\label{eq:b_eff}
\ee
where ${\cal N}(z,M)$ is the number of objects actually present in the
catalogue with redshift in the range $z,~z+dz$ and $M$ in the range
$M,~M+dM$, whose integral over $\ln M$ is ${\cal N}(z)$. In our
analysis of cluster correlations we will use for ${\cal N}(z)$ in
eq.(\ref{eq:xifund}) the observed one, while in the theoretical
calculation of the effective bias we will take the ${\cal N}(z,M)$
predicted by the model described below. This phenomenological approach
is self-consistent, in that our theoretical model for ${\cal N}(z,M)$
will be required to reproduce the observed cluster abundance and their
$\log N$--$\log S$ relation.

For the cluster population it is extremely reasonable to assume that
structures on a given mass scale are formed by the hierarchical
merging of smaller mass units; for this reason we can consider
clusters as being fully characterized at each redshift by the mass $M$
of their hosting dark matter haloes. In this way their comoving mass
function $\bar n(z,M)$ can be computed using an approach derived from
the Press-Schechter technique. Moreover, it is possible to adopt for the
monochromatic bias $b(M,z)$ the expression which holds for virialized
dark matter haloes (e.g. Mo \& White 1996; Catelan et al. 1998).
Recently, a number of authors have shown that the Press-Schechter
(1974) relation does not provide an accurate description of the halo
abundance both in the large and small-mass tails (e.g. Sheth \& Tormen
1999). Also, the simple Mo \& White (1996) bias formula has been shown
not to correctly reproduce the correlation of low mass haloes in
numerical simulations. Several alternative fits have been recently
proposed (Jing 1998; Porciani, Catelan \& Lacey 1999; Sheth \& Tormen
1999; Jing 1999). In this paper we adopt the relations recently
introduced by Sheth \& Tormen (1999), which have been shown to produce
an accurate fit of the distribution of the halo populations in the GIF
simulations (Kauffmann et al. 1999). They read
\be
\bar n(z,M) =  \sqrt{2 a A^2 \over \pi} \ 
{{3 H_0^2 \Omega_{\rm 0m}} \over {8\pi G}} \ 
{ \delta_c 
\over M  D_+(z) \sigma_M } \ 
\biggl[ 1 + \biggl( {{D_+(z) \sigma_M} \over {\sqrt{a}  \delta_c}}
\biggr)^{2p} \biggr] \ 
  \bigg| {d \ln \sigma_M 
\over d \ln M} \bigg| \exp \biggl[ -{a \delta_c^2  \over 
2 D_+^2(z) \sigma^2_M}
\biggr] \; 
\label{eq:ps2}
\ee  
and
\be 
b(M,z) =   1 + {1 \over \delta_c }
\biggl( {a \delta_c^2 \over \sigma_M^2 D_+^2(z)} - 1\biggr) 
+ {2 p  \over \delta_c }
\biggl( {1 \over { 1+[\sqrt{a} \delta_c / (\sigma_M D_+(z))]^{2p}}}
\biggr) \; .
\label{eq:b_mono2}
\ee 
Here $\sigma^2_M$ is the mass-variance on scale $M$, linearly
extrapolated to the present time ($z=0$), and $\delta_c$ the critical
linear overdensity for spherical collapse. Following Sheth \& Tormen
(1999), we adopt their best-fit parameters $a=0.707$, $p=0.3$ and
$A\approx 0.3222$, while the standard (Press \& Schechter and Mo \&
White) relations are recovered for $a=1$, $p=0$ and $A=1/2$.  Notice
that
\be
{\cal N}(z,M) = 4\pi r^2(z) {dr \over dz}
\left[1 + {H_0^2 \over c^2} ~\Omega_{0\cal R} ~r^2(z)\right]^{-1/2} 
{\bar n}(z,M) ~\phi(z,M) \;,
\ee
where $\phi(z,M)$ is the isotropic catalogue selection function, which
also accounts for the catalogue sky coverage, as detailed below.

The last ingredient entering in our computation of the correlation
function is the redshift evolution of the dark matter covariance
function $\xi_{\rm m}$. As in Matarrese et al. (1997) and Moscardini
et al. (1998) we use an accurate method, based on the Hamilton et
al. (1991) ansatz, to evolve $\xi_{\rm m}$ into the fully non-linear
regime. In particular, we use the fitting formula given by Peacock \&
Dodds (1996).

In order to predict the abundance and clustering of X-ray clusters in
the RASS1 Bright Sample we need to relate X-ray cluster fluxes into a
corresponding halo mass at each redshift. The given band flux $S$
corresponds to an X-ray luminosity $L_X=4\pi d_L^2 S$ in the same
band, where $d_L=(1+z) r(z)$ is the luminosity distance. To convert
$L_X$ into the total luminosity $L_{\rm bol}$ we perform band and
bolometric corrections by means of a Raymond-Smith code, where an
overall ICM metallicity of $0.3$ times solar is assumed. We translate
the cluster bolometric luminosity into a temperature, adopting the
empirical relation $T = {\cal A} \ L_{\rm bol}^{\cal B} \
(1+z)^{-\eta}$, where the temperature is expressed in keV and $L_{\rm
bol}$ is in units of $10^{44} h^{-2}$ erg s$^{-1}$. In the following
analysis we assume ${\cal A}=4.2$ and ${\cal B}=1/3$; these values
allow a good representation of the local data for temperatures larger
than $\approx 1$ keV (e.g. David et al. 1993; White, Jones \& Forman
1997; Markevitch 1998).  Analysing a catalogue of local compact
groups, Ponman et al. (1996) showed that at lower temperatures the
$L_{\rm bol}-T$ relation has a steeper slope (${\cal B}\approx
0.1$). For these reasons we prefer to fix a minimum value for the
temperature at $T=1$ keV. Moreover, even if observational data are
consistent with no evolution in the $L_{\rm bol}-T$ relation out to $z
\approx 0.4$ (Mushotzky \& Scharf 1997), a redshift evolution
described by the parameter $\eta$ has been introduced to reproduce the
observed $\log N$--$\log S$ relation (Rosati et al. 1998; De Grandi et
al. 1999) in the range $2 \times 10^{-14} \le S \le 2 \times 10^{-11}$
(see also Kitayama \& Suto 1997; Borgani et al. 1999). The values of
$\eta$ required for SCDM, $\tau$CDM, TCDM, OCDM and $\Lambda$CDM
models are reported in Table \ref{t:models}. A general discussion of
the effects of different choices of the parameters entering in this
method (e.g. the scatter in the $L_{\rm bol}-T$ relation) is presented
elsewhere (Moscardini et al. 1999).

Finally, with the standard assumption of virial isothermal gas
distribution and spherical collapse, it is possible to convert the
cluster temperature into the mass of the hosting dark matter halo,
namely (e.g. Eke, Cole \& Frenk 1996)
\be
T = {7.75 \over \beta_{\rm TM}} {\left(M\over {10^{15} h^{-1} 
M_\odot}\right)}^{2/3} 
(1+z) {\left( \Omega_{\rm 0m} \over \Omega_{\rm m}(z)\right) }^{1/3}
\left({\Delta_{\rm vir}(z) \over {178}}\right)^{1/3} \ .
\ee
The quantity $\Delta_{\rm vir}$ represents the mean density of the
virialized halo in units of the critical density at that redshift
(e.g. Bryan \& Norman 1998 for fitting formulas). We assume
$\beta_{\rm TM}=1.17$, which is in agreement with the results of
different hydrodynamical simulations (Bryan \& Norman 1998; Gheller,
Pantano \& Moscardini 1998; Frenk et al. 1999).

Once the relation between observed flux and halo mass at each redshift
is established we account for the RASS1 Bright Sample sky coverage
$\Omega_{\rm sky}(S)$ (see Figure 7 in De Grandi et al. 1999) by
simply setting $ 4 \pi \phi(z,M) = \Omega_{\rm sky}[S(z,M)]$.

\subsection{Results}

\begin{figure}
\centering  
\psfig{figure=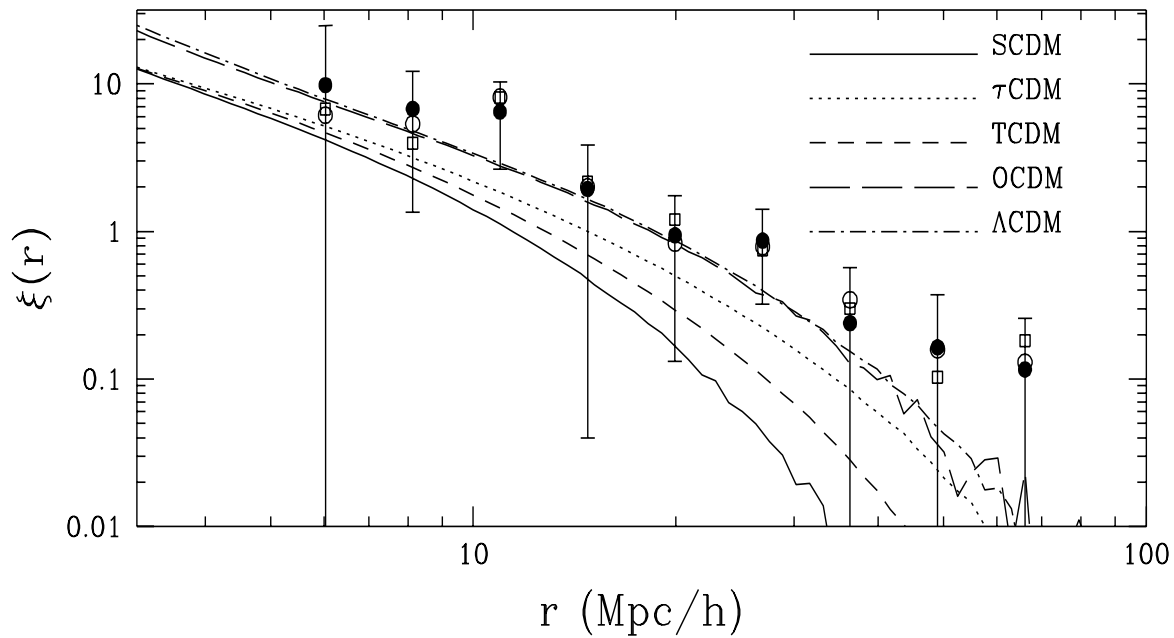,height=10.cm,width=16cm,angle=0}
\caption{Comparison of the observed spatial correlation for clusters in the
RASS1 Bright Sample (shown as in Figure~\ref{fi:xitot}) with the
predictions of the various theoretical models: SCDM (solid line),
$\tau$CDM (dotted line), TCDM (short-dashed line), OCDM (long-dashed
line) and $\Lambda$CDM (dotted-dashed line).  }
\label{fi:theor}
\end{figure}

In Figure~\ref{fi:theor} we compare our predictions for the RASS1
spatial correlation function in different cosmological models to the
observational data. All the EdS models here considered predict too
small an amplitude. Their correlation lengths are smaller than the
observational results: we find $r_0\simeq 11.5, 12.8, 14.8 \ h^{-1}$
Mpc for SCDM, TCDM and $\tau CDM$, respectively.  On the contrary,
both the OCDM and $\Lambda$CDM models are in much better agreement
with the data and their predictions are always inside the 1-$\sigma$
errorbars ($r_0\simeq 18.4, 18.6 \ h^{-1}$ Mpc, respectively). In
order to quantify the differences between the model predictions and
the observational data, we use again the maximum likelihood approach.
The minimum value for $S$ is obtained for the $\Lambda$CDM model. A
similar value, corresponding to $\Delta_S=0.7$, is obtained for
the OCDM model, while for the EdS models the resulting $S$ are much
larger: $\Delta_S=10.6, 19.3, 26.8$ for $\tau CDM$, TCDM and SCDM
models, respectively.

To evaluate what is the effect of neglecting the description of
clustering in the past light-cone, as usually done in previous
analyses, we estimate the cluster correlation function as
$\xi(r)=b^2_{\rm eff}(z_{\rm med})\xi_{\rm m}(r,z_{\rm med})$, where
$z_{\rm med}$ is the median redshift of the catalogue. For the RASS1
Bright Sample we have $z_{\rm med}\simeq 0.08$. The resulting values
of the correlation lengths obtained in this way are $r_0\simeq 13.5,
15.5, 18.2, 21.9, 22.4 \ h^{-1}$ Mpc, for SCDM, TCDM, $\tau CDM$, OCDM and
$\Lambda CDM$, respectively. They are typically 20 per cent higher
than the estimates obtained by our method. This difference is due to
the fact that in the past light-cone formalism the matter correlation
functions (and bias factors) are weighted by a factor ${\cal
N}(z)/r(z)$, for which the average value on the whole sample
corresponds neither to the median nor to the mode of the redshift
distribution. Actually, the presence of the comoving radial distance
$r(z)$ at the denominator tends to shift the ``effective redshift'' to
smaller values of $z$. As a consequence, the true cluster
correlations, which are indeed measured in our past light-cone,
have typically smaller amplitudes than those estimated at the median
redshift of the catalogue. Of course, the importance of this effect
becomes larger when deeper surveys are considered.
%
%--------------------------------------------------------
%--------------------------------------------------------
%

\begin{figure}
\centering  
\psfig{figure=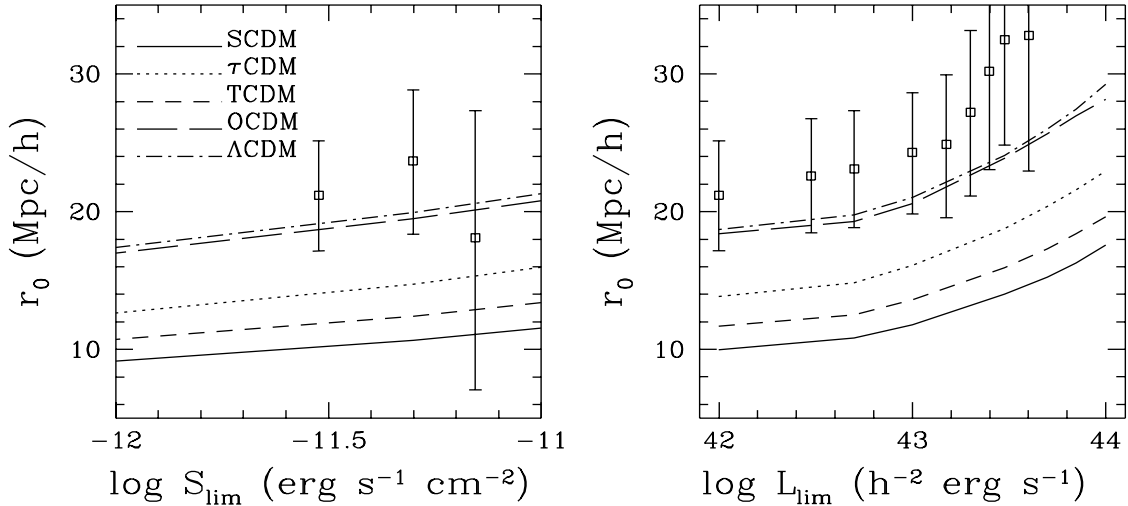,height=8.cm,width=16cm,angle=0}
\caption{The behaviour of the correlation length $r_0$
as a function of the limiting X-ray flux $S_{\rm lim}$ (left panel)
and luminosity $L_{\rm lim}$ (right panel). The open squares and
errorbars (95.4 per cent confidence level) refer to different
subsamples of RASS1 Bright Sample here analysed.  The prediction of
the various theoretical models are shown as in Figure~\ref{fi:theor}:
SCDM (solid line), $\tau$CDM (dotted line), TCDM (short-dashed line),
OCDM (long-dashed line) and $\Lambda$CDM (dotted-dashed line).  }
\label{fi:slim}
\end{figure}

In order to study the possible dependence of the clustering properties
of the X-ray clusters on the observational characteristics defining
the survey, we use our model to predict the values of the correlation
length $r_0$ in catalogues where we vary the limiting X-ray flux
$S_{\rm lim}$ or luminosity $L_{\rm lim}$ (both defined in the energy
band 0.5 -- 2 keV). Notice that this analysis can be related to the
study of the richness dependence of the cluster correlation
function. In fact, a change in the observational limits implies a
change in the expected mean intercluster separation $d_c$. Bahcall \&
West (1992) found that the Abell clusters data are consistent with a
linear relation $r_0=0.4 d_c$, while a milder dependence is resulting
from the analysis of the APM clusters (Croft et al. 1997). Our results
are shown in Figure~\ref{fi:slim}. All the cosmological models display
a similar trend: in the flux and luminosity intervals here considered,
the correlation length $r_0$ has a slow growth with $S_{\rm lim}$
(left panel), and a more marked one with $L_{\rm lim}$ (right
panel). For example, for OCDM and $\Lambda CDM$ the correlation length
changes from $r_0 \simeq 18$ to $r_0 \simeq 21 h^{-1}$ Mpc, when the
limiting flux varies from $S_{\rm lim}= 10^{-12}$ to $S_{\rm
lim}=10^{-11}$ erg s$^{-1}$ cm$^{-2}$, and from $r_0\simeq 18$ to
$r_0\simeq 30 h^{-1}$ Mpc, when the limiting luminosity varies from
$L_{\rm lim}= 10^{42}$ to $L_{\rm lim}= 10^{44} h^{-2}$ erg
s$^{-1}$. The values of $r_0$ for the EdS models have similar
variations but are always smaller. We can compare these predictions to
the results obtained by computing the two-point correlation function
in the RASS1 Bright Sample with the same cuts in X-ray flux or
luminosity. The estimates of $r_0$ obtained in this way are presented
in Table \ref{t:r0_lim} (where we also report the number of clusters
inside each subsample) and shown in the figure (open squares).  In
both panels the whole catalogue is represented by the square on the
right.  Because of the small number of clusters in these subsamples,
we prefer to fit the correlation function by fixing the value of the
slope $\gamma=1.8$; we find that our values of $r_0$ are only slightly
dependent on this assumption. The errorbars shown in the figure
correspond to an increase $\Delta_S = 4$ with respect to the minimum
value of $S$.  With the assumption that $\Delta_S$ is distributed as a
$\chi^2$ distribution with a single degree of freedom, this
corresponds to 95.4 per cent confidence level. By analysing the trend
with changing limiting flux, we find that the observed values of $r_0$
are almost constant even if $S_{\rm lim}$ changes by a factor larger
than 2. This result is in agreement with what Borgani, Plionis \&
Kolokotronis (1999) obtained for XBACs. We notice that OCDM and
$\Lambda$CDM are able to reproduce the amount of clustering shown by
the RASS1 Bright Sample, while all the EdS models strongly
underpredict the amplitude of the correlation function. The situation
is slightly different when we analyse luminosity-limited
catalogues. The RASS1 Bright Sample suggests a small increase of $r_0$
with $L_{\rm lim}$, even if the hypothesis of a constant correlation
length cannot be rejected.  This is consistent with a similar analysis
made by Abadi, Lambas \& Muriel (1998) on the XBACs catalogue. As
shown in the left panel of Figure~\ref{fi:slim}, our models always
tend to predict smaller correlations, even if the non-EdS models are
still marginally consistent with the observational data.

%
%--------------------------------------------------------
%--------------------------------------------------------
%
\begin{table}
\centering
\caption[]{The correlation length $r_0$ in flux-limited 
(left) and luminosity-limited (right) subsamples of the RASS1 Bright
Sample.  Column 1: the limiting X-ray flux $S_{\rm lim}$ (in units of
$10^{-11}$ erg s$^{-1}$ cm$^{-2}$) or luminosity $L_{\rm lim}$ (in
units of $10^{44} h^{-2}$ erg s$^{-1}$); Column 2: the number of
clusters in the subsample; Column 3: the correlation length $r_0$ (in
units of $h^{-1}$ Mpc) as obtained from the maximum likelihood
analysis with fixed slope $\gamma=1.8$ (errorbars correspond to 95.4
per cent confident level) }
\tabcolsep 4pt
\begin{tabular}{ccccccc} \\ \\ \hline \hline
$S_{\rm lim}$ & no. of clusters & $r_0$ &\hspace{2.truecm} & $L_{\rm lim}$ & 
no. of clusters & $r_0$ \\ \hline
0.3 & 126 & $21.2^{+4.0}_{-4.1}$ && 0.01& 126 & $21.2^{+4.0}_{-4.1}$\\
0.5 &  78 & $23.7^{+5.2}_{-5.4}$ && 0.03& 122 & $22.6^{+4.2}_{-4.2}$\\
0.7 &  41 & $18.1^{+9.3}_{-11.1}$ &&0.05& 118 & $23.1^{+4.3}_{-4.3}$\\
& & & &0.10 & 115 & $24.3^{+4.4}_{-4.5}$ \\
& & & &0.15 & 106 & $24.9^{+5.1}_{-5.4}$ \\
& & & &0.20 &  98 & $27.2^{+6.0}_{-6.1}$ \\
& & & &0.25 &  89 & $30.2^{+7.1}_{-7.2}$ \\
& & & &0.30 &  84 & $32.5^{+7.6}_{-7.7}$ \\
& & & &0.40 &  74 & $32.8^{+9.5}_{-9.9}$ \\\hline
\end{tabular}
\label{t:r0_lim}
\end{table}

\section{Conclusions} 

In this paper we have studied the two-point correlation function
$\xi(r)$ of a flux-limited sample of X-ray galaxy clusters, the RASS1
Bright Sample. These observational results have been used to test a
theoretical model predicting the clustering properties of X-ray
clusters in flux-limited surveys in the framework of different
cosmological scenarios. Our main results are:
\begin{itemize}
\item Assuming an Einstein-de Sitter model, 
$\xi(r)$ can be well fitted using the standard power-law relation
$\xi=(r/r_0)^{-\gamma}$, with $r_0= 21.5^{+3.4}_{-4.4} h^{-1}$ Mpc and
$\gamma=2.11^{+0.53}_{-0.56}$ (95.4 per cent confidence levels with
one fitting parameter). The values obtained in models with matter
density parameter $\Omega_{\rm 0m}=0.3$ are quite similar.
\item 
The amplitude of the correlation function is almost constant when the
RASS1 Bright Sample is analysed with different limiting fluxes in the
range $S_{\rm lim}=3-7 \times 10^{-12}$ erg s$^{-1}$ cm$^{-2}$ while
it displays a slightly increasing trend when computed in catalogues with an
increasing X-ray luminosity limit in the range $L_{\rm lim}=0.01-0.4
\times 10^{44}h^{-2}$ erg s$^{-1}$. 
\item
The comparison with the predictions of our theoretical models shows
that the Einstein-de Sitter models are unable to reproduce the
observational results obtained for the whole RASS1 Bright Sample. On
the contrary, good agreement is found for the models with matter
density parameter $\Omega_{\rm 0m}=0.3$, both with and without a
cosmological constant.
\item 
Our models are also able to reproduce the behaviour of $r_0$ with
$S_{\rm lim}$ and $L_{\rm lim}$, but the observed amount of clustering
is reproduced only by the open and $\Lambda$ models.
\end{itemize}

In conclusion, we believe that the method presented here leads to
robust predictions on the clustering of X-ray clusters; its future
application to new and deeper catalogues will allow to provide useful
constraints on the cosmological parameters.

\section*{Acknowledgments.}  
This work has been partially supported by Italian MURST, CNR and
ASI. We are grateful to Stefano Borgani, Enzo Branchini, Houjun Mo,
Ornella Pantano, Piero Rosati, Bepi Tormen and Elena Zucca for useful
discussions. We also thank the referee, C. Collins, for comments which
allowed us to improve the presentation of this paper.

\end{document}